\documentclass{article}
\usepackage{ijcai23}

\usepackage{times}
\usepackage{soul}
\usepackage{url}
\usepackage{amsfonts}
\usepackage[hidelinks]{hyperref}
\usepackage[utf8]{inputenc}
\usepackage[small]{caption}
\usepackage{graphicx}
\usepackage{amsmath}
\usepackage{amsthm}
\usepackage{booktabs}
\usepackage{algorithm}
\usepackage{algorithmic}
\usepackage[switch]{lineno}
\usepackage{bbding}
\usepackage{authblk}
\usepackage{pifont}

\title{NAS-FM: Neural Architecture Search for Tunable and Interpretable Sound Synthesis based on Frequency Modulation}

\author[1]{Zhen Ye}
\affil[1]{Hong Kong Baptist University}
\author[2]{Wei Xue$^*$}
\affil[2]{Hong Kong University of Science and Technology}
\author[3]{Xu Tan}
\author[4,2]{Qifeng Liu}
\author[2]{Yike Guo\thanks{Corresponding authors: Wei Xue \{\small weixue@ust.hk\}, Yike Guo\{\small yikeguo@ust.hk\}} }
\affil[3]{Microsoft Research Asia}
\affil[4]{Hong Kong Institute of Science \& Innovation, Chinese Academy of Sciences}

\usepackage[bottom]{footmisc}

\begin{document}

\maketitle

\begin{abstract}
Developing digital sound synthesizers is crucial to the music industry as it provides a low-cost way to produce high-quality sounds with rich timbres. Existing traditional synthesizers often require substantial expertise to determine the overall framework of a synthesizer and the parameters of submodules. Since expert knowledge is hard to acquire,  it hinders the flexibility to quickly design  and tune digital synthesizers for diverse sounds. In this paper, we propose ``NAS-FM'', which adopts neural architecture search (NAS) to build a differentiable frequency modulation (FM) synthesizer. Tunable synthesizers with interpretable controls can be developed automatically from sounds without any prior expert knowledge and manual operating costs. In detail, we train a supernet with a specifically designed search space, including predicting the envelopes of carriers and modulators with different frequency ratios. An evolutionary search algorithm with adaptive oscillator size is then developed to find the optimal relationship between oscillators and the frequency ratio of FM. Extensive experiments on recordings of different instrument sounds show that our algorithm can build a synthesizer fully automatically, achieving better results than    handcrafted synthesizers. Audio samples are available at \href{https://nas-fm.github.io/} {https://nas-fm.github.io/}.
\end{abstract}

\section{Introduction}
Creating and rendering music has become increasingly convenient with the help of digital sound synthesizers which simulate the timbre of real instruments that are potentially expensive and rare. Lots of sound synthesizers are designed whose commercial values are widely recognized, while the complexity of the synthesizers also increases dramatically.

\begin{table}[!t]
\resizebox{\linewidth}{!}{
\begin{tabular}{c|ccc}
\hline
Synthesizer                                                          & Tunable & Interpretable & \begin{tabular}[c]{@{}c@{}}Automatic\\ Design\end{tabular} \\ \hline
\begin{tabular}[c]{@{}c@{}}Digital\\ Synthesizer\end{tabular}        & \Checkmark              & \Checkmark                                                             & \XSolidBrush             \\ \hline
\begin{tabular}[c]{@{}c@{}}Neural Network\\ Synthesizer\end{tabular} & \XSolidBrush             & \XSolidBrush                                                             & \Checkmark             \\ \hline

Our NAS-FM                                                        & \Checkmark              & \Checkmark                                                             & \Checkmark             \\ \hline

\end{tabular}}
\caption{Comparison between the proposed NAS-FM with other strategies for digital audio synthesis.}
\end{table}

Instead of carefully designing delicate structures of highly diversified instruments, our target is to design a general framework which can flexibly construct digital synthesizers simply from recordings and allow further tuning of the timbres in a controllable and interpretable way.

Early approaches to sound synthesis are parametric, designing a set of components (e.g., oscillators and modulators) based on digital signal processing (DSP), and ultimately creating complex timbres with specific structures combining these components. Typical works include subtractive synthesis \cite{huovilainen2005new}, additive synthesis \cite{serra1990spectral}, frequency modulation (FM) \cite{chowning1973synthesis} and wavetable synthesis \cite{bristow1996wavetable}, among which FM-based methods are widely used in sound synthesis due to their flexibility in tuning timbre with only a few parameters. 

While satisfying sounds can be produced, designing the parametric structure of the synthesizer as well as parameter values requires considerable expertise. This is due to the non-linear interactions between synthesizer parameters, as well as the wide range of possible values for each parameter. Achieving a desired sound often involves a process of iterative adjustments and fine-tuning, which can be time-consuming and require a deep understanding of digital signal processing techniques. Although some efforts have been made to determine the parameters of FM synthesizer based on estimation theory \cite{justice1979analytic}, genetic algorithm \cite{horner1998nested},    LSTM-based \cite{yee2018automatic}, VAE-based \cite{le2021improving} and dilated CNN based \cite{chen2022sound2synth} methods, these methods are limited by   static spectra or an audio clip conditioned on an entire ADSR envelope under a specific pitch value and a fixed duration length. Thus, these methods cannot be applied to the audio    with continuously varying pitch and loudness on dynamic spectra.  

Neural synthesis methods have been developed recently to learn a deep neural network to produce audio  in the data-driven scheme and achieved impressive results in terms of audio fidelity. However, for fully-neural generative models such as WaveGlow \cite{prenger2019waveglow}, and HiFi-GAN \cite{kong2020hifi}, sufficient data is usually required to train a complex model with  non-interpretable parameters, which greatly limits the controllability of the synthesizer as required by the music industry, specifically when the generated non-perfect sound needs tuning. Although the harmonic-plus-noise model-based differential DSP (DDSP) is integrated into the network  to improve controllability \cite{engel2020ddsp}, it only allows the transfer of timbres between different instruments rather than explicit adjustment. FM-based neural synthesizer  DDX7 \cite{caspe2022ddx7} is then developed.  However,  this method can only learn the time-varying parameters of the submodules under the assumption that the overall structure has been manually defined.

The large reliance on expert knowledge and extensive time cost greatly limit the feasibility of building digital synthesizers for general instruments. Moreover, the handcrafted framework may make the resulting synthesizer suboptimal in modelling the real instrument. This paper proposes NAS-FM, a neural architecture search (NAS) based FM synthesizer. It can be seen as an effort to build interpretable neural generative models. A key of the proposed method is NAS. With a carefully designed search procedure, different structures and oscillator sizes are included in a universal search space of the supernet, and an evolutionary search algorithm is developed to find the optimal structure between oscillators and the frequency ratio of FM. The advantages of the proposed NAS-FM are described below:
\begin{itemize}
\item The audio synthesizer can be built based on recordings without any expert knowledge, largely simplifying the pipeline and reducing the cost of audio synthesizer construction. It is also possible to quickly build new variants of the target sound with flexible adjustments;
\item It returns a tunable and interpretable conventional FM-based interface with a few parameters, making it easily embedded into existing audio workstations;
\item Extensive experiments on recordings of different instruments demonstrate that synthesizers fully automatically built by the proposed NAS-FM can achieve better results to carefully handcrafted synthesizers. 
\end{itemize}

\section{Related Work}
\subsection{Sound Synthesis}
Sound synthesis contains  digital signal processing (DSP) methods and neural network methods. DSP methods have been integrated into the  digital audio workstation used by musicians. More specifically, DSP methods start from several simple waves such as sine, sawtooth, square and triangle generated by an oscillator.
The additive synthesizer \cite{serra1990spectral} generates new sounds by adding various simple waves. The Subtractive synthesizer \cite{huovilainen2005new} filters a simple wave from white noises. FM synthesizers \cite{chowning1973synthesis} rely on simple waves to modulate frequency to create complex timbre. Wavetable synthesis 
manipulate a collection of short samples to design new sounds.
These traditional methods need users to determine the configuration manually for a given sound.

Neural network synthesis models adopt the deep neural network to learn the  mapping function  between audio and given input, for instance, pitch and loudness.  The early exploration begins with auto-regressive models such as WaveRNN \cite{kalchbrenner2018efficient} and SampleRNN \cite{mehri2016samplernn}. The following works \cite{engel2019gansynth} \cite{kong2020diffwave} are based on various generative models to further improve the quality of synthesis sound. However, the above methods may lead to glitch problems because of the lack of phase continuity. Therefore, DDSP \cite{engel2020ddsp} rely on the harmonic-plus-noise model to keep the phase continuous and also make the sound can be directly controlled by pitch and loudness. While these methods can be optimized automatically with the help of gradient descent to obtain the model, there are few control factors to help users manipulate the synthesized result directly. Therefore, our approach aims to introduce the FM synthesizer with controllable factors to help the user interact with the synthesized audio.

\subsection{FM Parameter Estimation}
FM parameter estimation also called FM matching is adopted to determine the configuration of an   FM synthesizer. The early approach considers this problem as a searching problem that uses the genetic algorithm(GA) \cite{jh1975adaptation} or its variants to find a best-fit configuration in a specific searching space. Horner employs GA solving a sound matching problem for different FM algorithms such as Formant FM \cite{horner1993machine}, Double FM \cite{horner1996double}, Nested FM and Feedback FM \cite{horner1998nested}.  These methods can achieve very close re-synthesizing results with a static target spectrum when selecting an  appropriate FM algorithm as  prior. The following methods leverage an open-source FM synthesizer Dexed synthesizer \cite{Gauthier2013dexed} to construct a large number of pair data between presets\footnote{Preset means a  full FM configuration for specific sound designed by users}collected on the Internet and synthesized audio clip generated by Dexed. By reversing this process, these works employ LSTM \cite{yee2018automatic}, VAE \cite{le2021improving} and dilated CNN \cite{chen2022sound2synth} to estimate the preset. Since the synthesized sound is generated by pressing a note with specific velocity, duration and pitch using Dexed synthesizer, the result of the model for a realistic sound audio clip without any prior is unpredictable. DDX7 \cite{caspe2022ddx7} predict the  envelopes of oscillators using the widely-used TCN \cite{oord2016wavenet} decoder with algorithm and frequency ratios as prior.
Therefore, our method is designed to construct an FM synthesizer without prior knowledge. 

\subsection{Neural Architecture Search}
Neural Architecture Search (NAS) can automatically find the best neural architecture for a specific task. Many works focus on computer vision  \cite{liu2018darts}or natural language processing \cite{xu2021bert} tasks.  More recently, one-shot NAS \cite{guo2020single} \cite{bender2018understanding} train a shared supernet once which includes all candidate architectures. Then, the supernet  is used as an estimator to evaluate every possible   architecture in the search space. This method has been widely used on various applications to determine a good structure, for example, image classification \cite{guo2020single}, object detection  \cite{liang2021opanas}, 3d scene understanding \cite{tang2020searching}, BERT compression \cite{xu2022analyzing}. These methods introduce neural architecture search to each specific task which gets a better model architecture than manual design. Although  NAS has been widely used in lots of areas, the applications are mainly focused on neural networks. Actually, we are the first to bring   NAS to fully automatically build  the sound synthesizer.

\section{Review of FM Synthesizer }
\subsection{Basics of FM} \label{Frequency Modulation}

FM was originally proposed in \cite{chowning1973synthesis} for sound synthesis and different timbres are produced by controlling a set of parameters. With two sound sources, the modulator oscillator $\sin (2\pi {f_m}t)$ and the carrier oscillator $\sin (2\pi {f_c}t)$, FM basically generates a time-domain signal $y(t)$:
\begin{equation}
y(t) =a(t) \sin (2\pi {f_c}t + I\sin (2\pi {f_m}t)),
\label{fm}
\end{equation}
where $f_c$ and $f_m$ are the carrier frequency and modulation frequency respectively, $I$ is the modulation index, and $a(t)$ is the amplitude envelope of the carrier. $y(t)$ can be further decomposed by using Bessel functions of the first kind as
\begin{equation}
y(t) =a(t) \sum\limits_{n=-\infty }^{n=+\infty } {{J_n}(I)\sin (2\pi ({f_c} + n{f_m}) \cdot t)} 
\label{eq2}
\end{equation}
which shows that the sidebands of $y(t)$ distribute evenly around $f_c$ with spacing as $f_m$, and the spectra is harmonic when the frequency ratio $r\in\mathbb{Q}$  where $r=f_c/f_m$. ${J_n}(I)$ is a Bessel function of the modulation index $I$, and dynamic spectra can be generated if the $I$ becomes a time-variant function $I(t)$.
 
\begin{figure}[!t]
      \centering
      \includegraphics[scale=0.4]{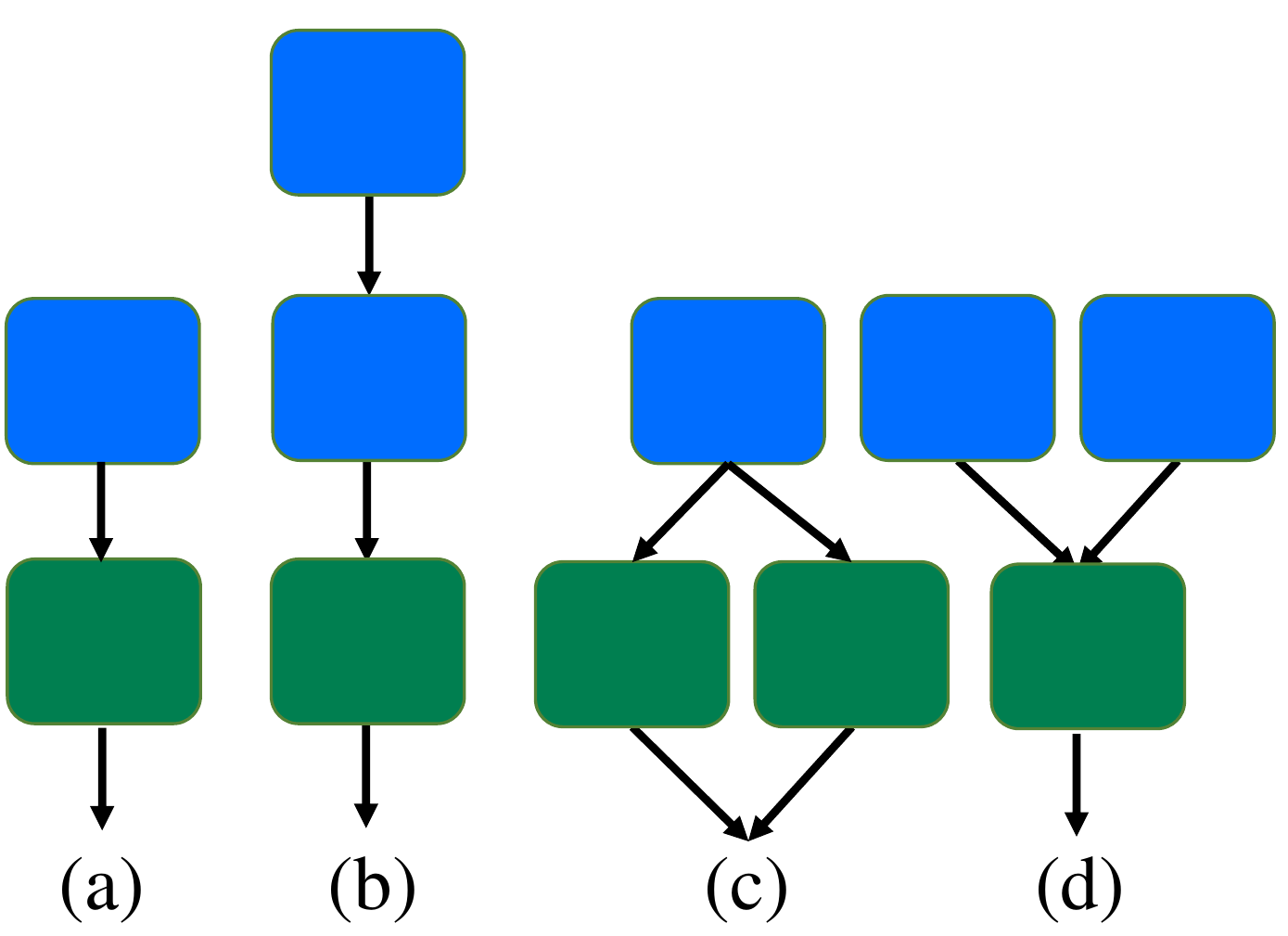}
      \caption{Different FM algorithms: (a) Single FM; (b) Nested FM; (c) Formant FM; (d) Double FM. The green box refers to the carrier and the blue box refers to the modulator}
      \label{4fm}
\end{figure}

\begin{figure}[!t]
      \centering
      \includegraphics[scale=0.32]{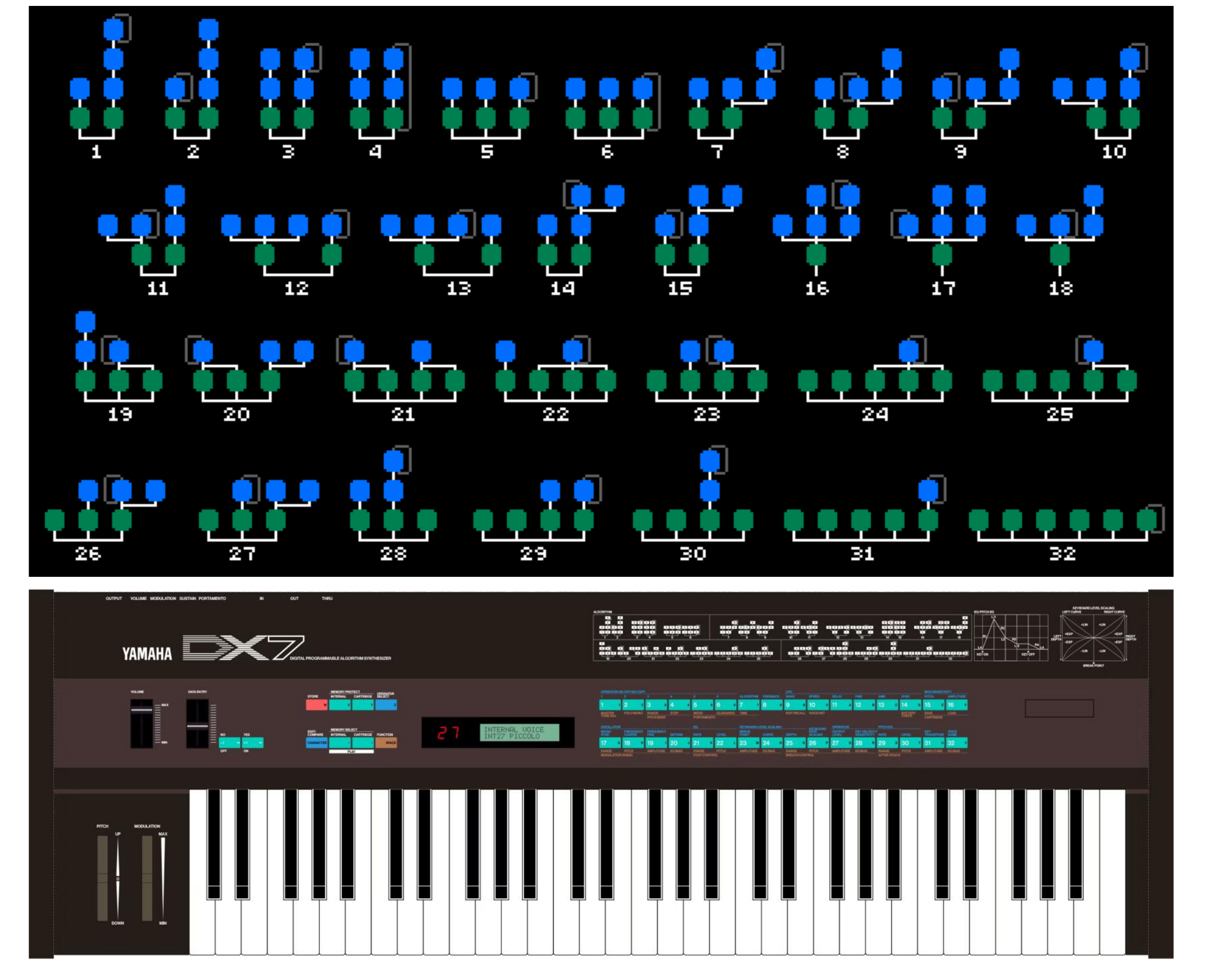}
      \caption{Digital   synthesizers in the commercial product YAMAHA DX7. The lower part is the user interface of a synthesizer. The upper part shows the FM algorithms for audio synthesis, with green and blue boxes denoting carriers and modulators, respectively.}
      \label{dx7}
\end{figure}
\subsection{FM algorithms}
The \eqref{fm} explains the basic module of the FM synthesizer to generate sounds. More diversified and expressive FM synthesizers can be further developed by designing complicated topologies, which are called ``FM algorithms'', on connecting the carrier and modulator oscillators. 

Typical FM algorithms are shown in Fig.~\ref{4fm}. Fig.~\ref{4fm}(a) denotes the single FM expressed by \eqref{fm}. The Nested FM \cite{justice1979analytic}, formant FM \cite{horner1993machine}, and double FM \cite{schottstaedt1977simulation} are illustrated in Fig.~\ref{4fm}(b)-(d), whose outputs are calculated by
\begin{align}
y(t) = &a(t)\sin ( 2\pi {f_c}t + {I_{m1}(t)}\sin [2\pi {f_{m1}}t \label{eq3}\nonumber\\
		&  + {I_{m2}(t)}\sin (2\pi {f_{m2}}t)]),\\
y(t) = &{a_1}(t)\sin [2\pi {f_{c1}}t + {I_m(t)}\sin (2\pi {f_m}t)] \nonumber\\
	   & +{a_2}(t)\sin [2\pi {f_{c2}}t + {I_m(t)}\sin (2\pi {f_m}t)],
\end{align}
and 
\begin{align}
y(t) = &a(t)\sin [2\pi {f_c}t + {I_{m1}(t)}\sin (2\pi {f_{m1}}t) \nonumber\\
&+ {I_{m2}(t)}\sin (2\pi {f_{m2}}t)],
\end{align} respectively. These FM algorithms produce different timbres, and as shown in Fig.~\ref{dx7}\footnote{We do not consider the  feedback FM in our work as the same reason in \cite{caspe2022ddx7}}, by selecting different FM algorithms. 

To briefly summarize, a set of parameters controls the FM output, which can be time-variant or fixed. In our work,  time-variant parameters include the oscillators' envelopes, i.e., $a(t)$ and $I(t)$, which are determined by the time-variant input, such as $f_0$ and loudness. Fixed parameters are the chosen FM algorithm and the frequency ratio of each oscillator, which need to be determined by the expected timbre.

\begin{figure}[!t]
    \centering
      \includegraphics[scale=0.5]{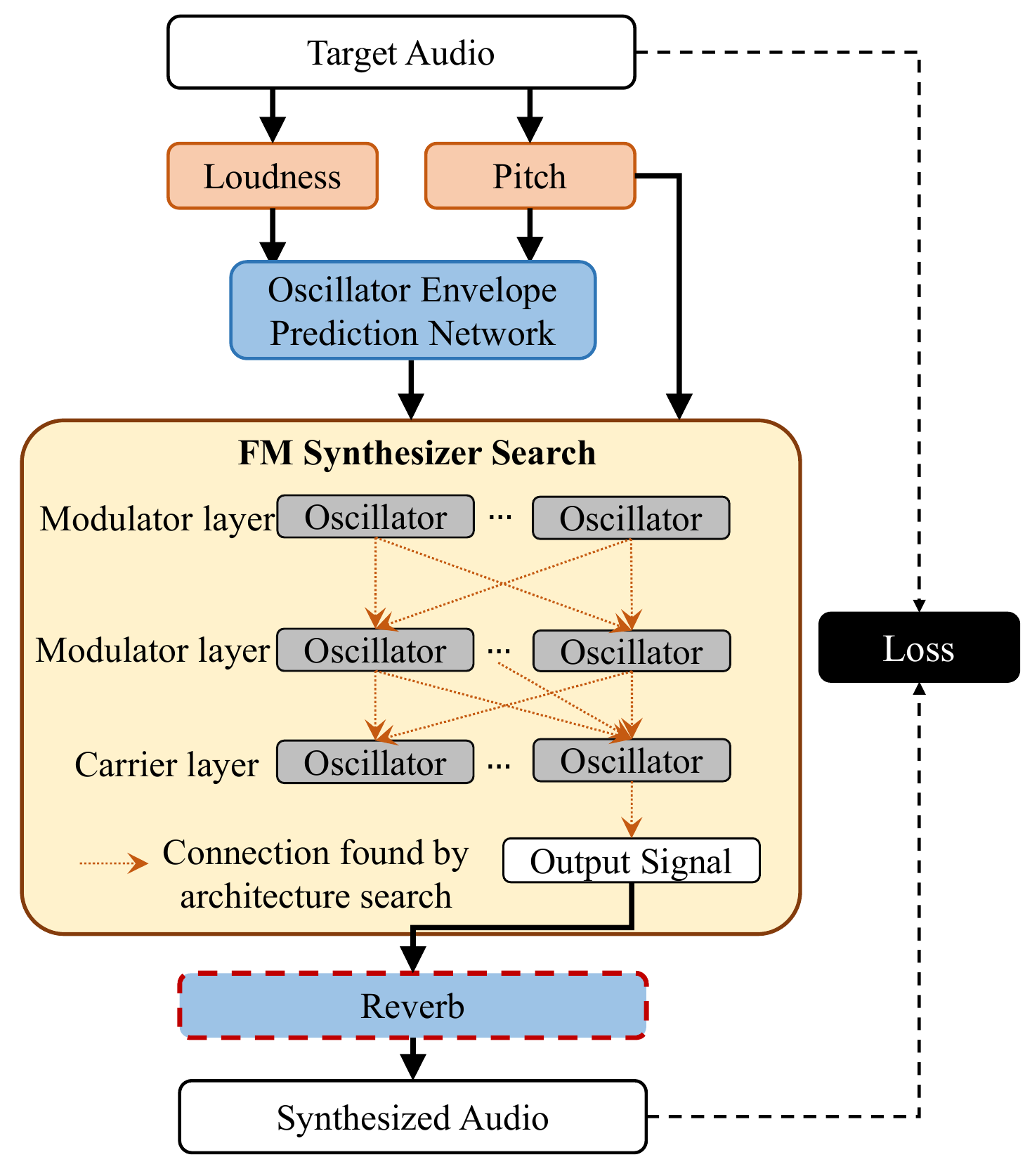}
      \caption{ The overall architecture of NAS-FM. The learnable Reverb module is optional depending on whether the real room effect should be simulated for real-world audios.}  
      \label{overall}
\end{figure}

\section{Proposed NAS-FM synthesizer}
In this section, we aim to fully automatize the designing of an FM synthesizer in a data-driven manner, thus eliminating the reliance on expertise and labour to tune the timbres. 

A NAS-FM synthesizer is proposed, with an overall framework shown in Fig.~\ref{overall} on the next page. For input audio, the pitch and loudness are extracted, and these two features are fed into an oscillator envelope prediction network to estimate the envelopes of the oscillators. We develop architecture search methods to determine the optimal FM algorithm, and given the FM algorithm and oscillator envelopes, sounds with a specific timbre can be produced. Depending on whether the sound is from the real environment, a learnable reverb module \cite{engel2020ddsp} can be optionally used to simulate the room effects. The framework is optimized in an auto-encoder setting, i.e., seeking to recover the original signal in the final output.

Besides the FM algorithm search, the framework is similar to the framework in DDX7 \cite{caspe2022ddx7}, which estimates pitch and loudness by CREPE \cite{kim2018crepe} and A-weighting loudness \cite{moore1997model}, designs the oscillator envelope prediction network as a temporal convolutional network (TCN) \cite{bai2018empirical}. However, DDX7 requires  a prior FM configuration designed by experts. By adopting NAS, the proposed framework designs the FM synthesizer in the fully-automated data-driven pipeline and also returns a tunable and interpretable interface used by musicians. In the following, details of the proposed NAS-FM will be introduced, which include a) converting the FM algorithm to a graph, b) designing the search space, c) training the supernet, and d) selecting the FM algorithm.

\subsection{Directed Acyclic Graph of FM Algorithm}
To facilitate discussion in this section, we convert an FM synthesizer, with examples shown in Fig.~\ref{4fm}, to a directed acyclic graph (DAG) \cite{pham2018efficient} with an ordered sequence of $N$ nodes, where $N$ represents the number of oscillators. Each node $x^{(i)}$  refers to the oscillator's output. Each directed edge $(i,j)$ indicates whether node $x^{(j)}$ is the modulating node of $x^{(i)}$. The topology of the graph is associated with the FM algorithm. Thus, an intermediate node $x^{(i)}$ can be expressed as
\begin{equation}
{x^{(i)}(t)} = {a_i}(t)\sin (2\pi {f_i}t + \sum\limits_{j \in \mathbb{M}} {{x^{(j)}(t)}} ),
\label{eq6}
\end{equation}
where $\mathbb{M}$ is a set of oscillators modulating node $x^{(i)}$. When $M$ is empty, $x^{(i)}$ outputs standard sine wave. The output of FM is calculated through the sum of carrier nodes
\begin{equation}
y(t) = \sum\limits_{i \in \mathbb{C}} {{x^{(i)}}(t)},
\label{fmoutput}
 \end{equation}
where $\mathbb{C}$ is the set of carriers.

\subsection{Search Space Design}
There are a huge number of possible configurations for the FM synthesizer. By converting the FM algorithm to the DAG, the principles of NAS can be applied to design the FM algorithm. In NAS for neural networks, all possible architectures of the network, which span the ``search space'', can be represented by a general DAG, with each candidate architecture as a sub-graph. Similarly, we design the search space for FM algorithm here.

\begin{figure}[!t]
      \centering
      \includegraphics[scale=0.45]{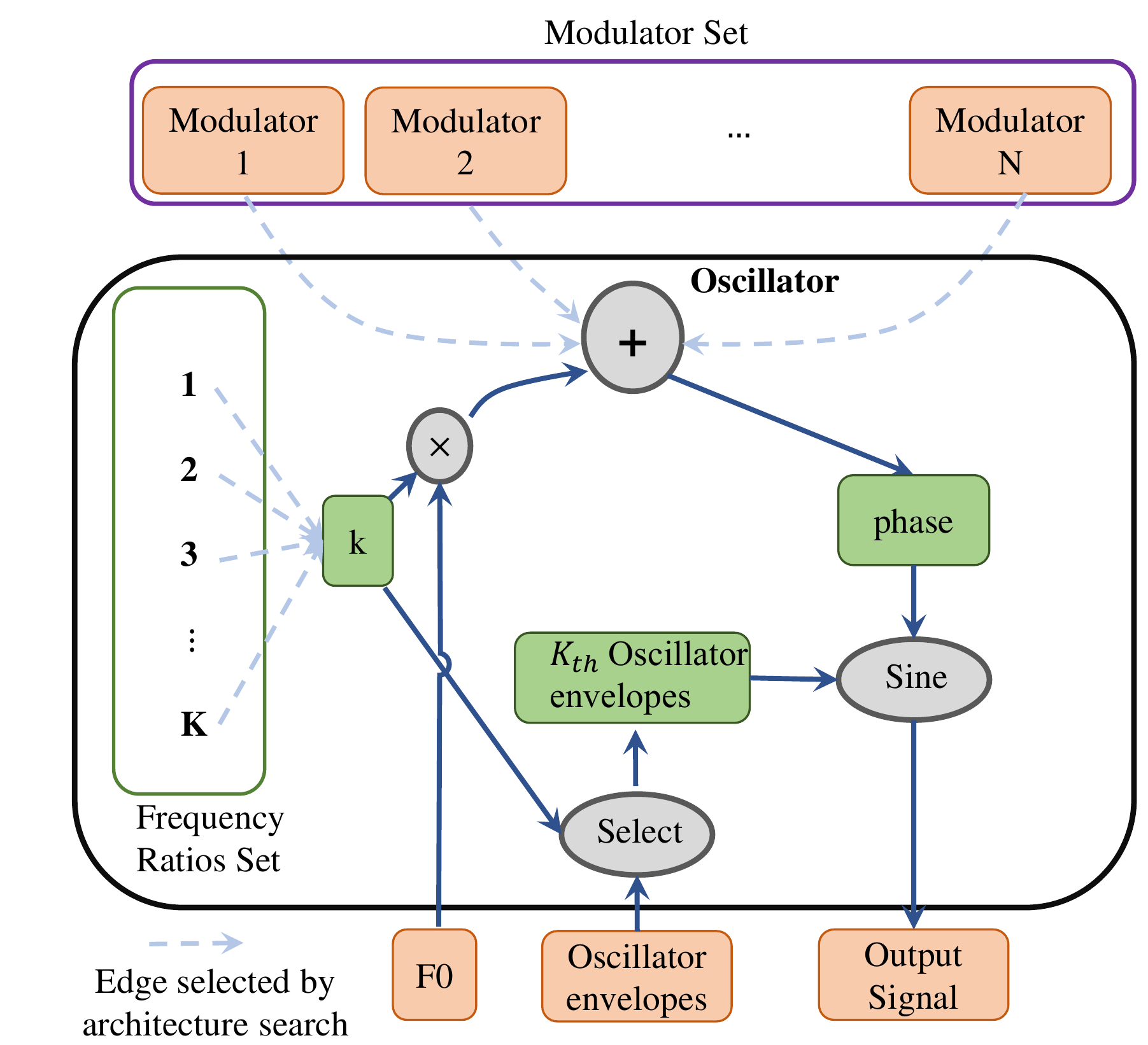}
      \caption{ The oscillator in NAS-FM. The output of an oscillator conditioning on $F0$ depends on the choice of  frequency ratio, envelope and modulator. The choices are determined by NAS.}
      \label{local}
\end{figure}

We visualize the operation in \eqref{eq6} in Fig.~\ref{local}, where two issues should be solved to define an oscillator: a) what is the frequency ratio? The frequency ratio is defined originally in \eqref{eq2}, and is the ratio between $f_i$ in \eqref{eq6} and fundamental frequency $F_0$ here; b) which modulators will be connected to the current oscillator? As depicted in Fig.~\ref{local}, we define the frequency ratio set $\mathbb{FR}=\{1,2,3,..,K\}$ which consists of $K$ integers, and assumes that there are $N$ candidates in the modulator set $\mathbb{M}$. Actually, the interval in the frequency ratio set can be reduced to 0.1 or even smaller, to 0.01 instead of 1, for a more refined exploration.

Traversing the original search space is impossible considering the exponential number of possibilities for oscillator connections. 
Now let us analyze the design of the search space. Previous works adopted evolutionary search on frequency ratio set under a fixed FM algorithm such as double FM \cite{horner1996double} and nested FM \cite{horner1998nested}. Therefore, the challenge is how to design an appropriate space including various FM algorithms, instead of enumerating all possible FM algorithms. In \cite{guo2020single} and \cite{xu2022analyzing}, the weight sharing is proposed to reduce the search space, which forces sub-graph candidates to have the same weights in the graph nodes commonly shared by different candidates. This also makes it possible to directly train a supernet including all possible candidates. Here we propose a novel envelope-sharing strategy for FM synthesizers to make all FM configuration candidates trained  in the supernet.

Specifically, we construct a search space as shown in Fig.~\ref{overall}. The search space can be divided into a carrier layer and several modulator layers. We set the oscillator  at the same layer with the same frequency ratios sharing the same envelope. Due to the envelope-sharing strategy, 
There are following rules in our search space.
(1) A  oscillator in a certain layer can only be modulated by the upper layer; (2) The sum of the output of selected oscillators at the carrier layer forms the final signal;(3) A  oscillator is discarded when there is no connection with other oscillators or final output. 
In addition, according to our experience, we find two modulator layers are enough. The number of candidate oscillators in a layer depends on the  oscillator number in the expected FM synthesizer. 

\subsection{Supernet Training with Proxy Oscillator}
Although the proposed envelope-sharing strategy makes all possible FM configurations can be trained using a supernet that includes all candidates, we still encounter the challenge of a large search space. The huge search space makes it hard to evaluate each FM configuration in the supernet during training. 
To further improve the training efficiency, we propose to use the ``proxy oscillator'' as the proxy for all oscillators in each layer to determine the envelopes of the oscillators. 

Specifically, during supernet training, a fixed Nested FM in \eqref{eq3} with a carrier and two modulators are chosen, which has three layers in accordance with the settings in the Sec.~4.2. For each oscillator, uniform sampling \cite{guo2020single} is adopted to determine the frequency ratio. With these two configurations fixed, the envelopes of the oscillators are learned, and after training, the learned envelopes are shared within the same layer for a complicated search space. Utilizing the technique, we can expand the width of the search space flexibly.

\subsection{FM Algorithm Selection}
After the supernet training and search space design, we use the evolutionary algorithm to conduct FM algorithm selection. Specifically, we put all $N$ candidate oscillators $ol$ in order. a certain FM configuration is encoded as an individual   can be formulated as
\begin{equation}
\{ {f_{o{l_1}}},{f_{o{l_2}}},...,{f_{o{l_N}}},{l_{o{l_1}}}{l_{o{l_2}}},...,{l_{o{l_N}}}\} 
\end{equation}
where $ol_i$ is the $i_{th}$ oscillator,
 corresponding ${f_{o{l_i}}}$ and ${l_{o{l_i}}}$ indicate the selected frequency ratio and connection relationship. The connection relationship is the relation between the lower layer.  If the oscillator belongs to the carrier, the lower layer is the output signal.  If the connection relationship of this oscillator is none means it is discarded. 
Firstly, a random population is initialized within the initial space.  Secondly, we evaluate the fitness score of the generated  individuals and select the top individuals. The fitness function will be introduced in the experiment section. Thirdly, crossover and mutation are used to generate new individuals to update the population, until meeting the stopping criterion. 

\section{Experiments}
To evaluate the proposed NAS-FM approach which aims to automatically learn the FM synthesizer that is applicable to the music industry, we aim to answer the following questions: a) Given sound recordings, can the FM synthesizers learned by the proposed method be comparable to the manually designed synthesizers? b) We fuse different FM algorithms into one universal search space, is this strategy better than separately searching the best configuration for each FM algorithm? c) Can we controllably tune the timbre of the produced sounds or create new instrument timbres  by modifying the parameters of the learned FM synthesizer?

\subsection{Dataset}
We conduct experiments on the benchmark URMP dataset \cite{li2018creating}. Sound recordings of three real instruments, which are violin, flute, and trumpet, are chosen and the ``optimal'' manually designed FM algorithms of these instruments are given as in \cite{caspe2022ddx7}. The three instruments are also the most typical instrument of the strings, woodwinds, and brass. Each instrument recording is divided into 4-second segments, and silent segments are discarded. The loudness and pitch are estimated by the A-weighting loudness \cite{moore1997model} and CREPE \cite{kim2018crepe} methods. Each audio clip is resampled to the 16kHz sampling rate and analyzed with a frame size of 2048 and a hop size of 64, yielding 1000 frames. We split the dataset into train, validation, and test sets with proportions of 0.75, 0.125, and 0.125, respectively.

\subsection{Experimental Setup}
For training,  a supernet  contains $c*m*m$ paths where $c$ is the number of frequency ratios of the carrier, and $m$ is the number of frequency ratios in modulators. We set $c$ as 15 and $m$ as 5. The model adopts the stack TCN architecture \cite{bai2018empirical} as a sequence-to-sequence model to predict oscillator envelope through pitch and loudness. We first stack 4 TCN architecture to extract a hidden temporal feature. Then, we construct $(c+m+m)$ TCN architectures with different weights to predict the envelope of the corresponding oscillator from the hidden feature. In fact, other sequence-to-sequence models can also be employed to model the temporal relationship in our pipeline. In addition, uniform sampling of oscillators on each layer with different ratios is adopted. We use Adam optimizer with an initial learning rate of 3e-4. For regularize, we set the maximum value of the oscillator for the carrier and modulator as 1 and 2, respectively.
The exponential decay strategy  with a decreasing  factor of 0.98 every 10k steps  is adopted for the learning rate
.The whole training steps are 500k, and the batch size is 16.  

For searching, we use the evolutionary algorithm with a population size of $P$, crossover size of $P/2$, and mutation size of $P/2$ with mutation probability and max iterations of $T$ depending on the search space size. In addition, since we use the fixed number of the candidate oscillator, the discarded  candidate oscillator with different frequency ratios will cause the same fitness score. Therefore, we force the generated candidates with a unique fitness score to be legal.

After searching, we directly extract the weight of the sub-model with the best fitness score   from the supernet as   our final model. Actually, fine-tuning the model or training  from scratch with the searched FM configuration   may further improve the performance. However, our aim is not to focus on the resynthesis performance   but to get a controllable synthesizer   close to the target sound so that musicians can further utilize   it to tweak the sound  or do other more interesting things such as sound morphing  and sound interpolation. 

\subsection{Evaluation Metric}

We use Fréchet Audio Distance (FAD) \cite{kilgour2018fr} as the evaluation metric to measure the distance between real and generated sound.  This method extracts the embedding  of the audio  using a pre-trained VGG-like model. The distance is calculated as follows:
\begin{equation}
FAD= {\left\| {{\mu _r} - {\mu _g}} \right\|^2} + tr({\Sigma _r} + {\Sigma _g} - 2\sqrt {{\Sigma _r}{\Sigma _g}} )
\end{equation}
where $r$ and $g$ are the real audio and generated audio respectively, ${\mu _r}$ and ${\mu _g}$ are the mean vector of embedding and ${\Sigma _r}$ and ${\Sigma _g}$ are the covariances of embedding. 
A smaller Fréchet Audio Distance indicates a higher level of similarity between the distributions of real and generated data.  In our experiment, we calculate the Fréchet Audio Distance between the    real data and synthesize  validation data as the fitness score during searching.  And calculating the FAD between the synthesized test data as the final evaluation result.

\begin{table}
    \centering
    \begin{tabular}{lrrr}
        \toprule
        
          & \multicolumn{3}{ c}{Fréchet Audio Distance (↓) } \\
        \midrule
        Model  &Flute &Violin &Trumpet \\
        \midrule
        Test Data     &   1.180      & 0.308  &   0.554  \\
                    
        DDX7   &  7.841      & 3.497  &  4.442  \\
                 
        NAS-FM &  \textbf{7.077}     &    \textbf{3.255}   &     \textbf{3.384}   \\
               
        \bottomrule
    \end{tabular}
      \caption{FAD of resynthesis results using 6 oscillators}
    \label{tdx7}
\end{table}

\subsection{Comparison with Manually-Designed FM Synthesizers}
In this part, we are interested to know if our method could achieve comparable results  to  manually designed FM synthesizers. The baseline is DDX7 \cite{caspe2022ddx7} which the author retrieves on the web manually to find the patch with the most similar sound to the target sound. The patches they find have six oscillators, and we adopt the same oscillator number with them. Since the authors did not open source their evaluation code, we  train  the model following their method ten times with a random seed and took the best one as their result. 

In our method, we use a $3*3$ candidate oscillator search space forcing three oscillators discarded to ensure six oscillators. During the search, we follow the above procedure and set the population size as 1000 and max iterations as 50. Results are shown in Table~ \ref{tdx7}. The Test Data line means the Fréchet Audio Distance between the  entire real data from the real test data.  We can see that our NAS-FM outperforms the hand-designed DDX7 baseline across all musical instrument recordings. The results show that our NAS-FM can search for more comparative FM configurations than manually designed FM synthesizers.

\begin{table}
    \centering
    \begin{tabular}{lrrr}
        \toprule
        
          & \multicolumn{3}{ c}{Fréchet Audio Distance (↓) } \\
        \midrule
        Model  &Flute &Violin &Trumpet \\
        \midrule

        Nested FM   &  12.75    & \textbf{6.02}& \textbf{8.24}  \\
                 
        Formant FM &  \ 14.48    &  7.56      &     8.68    \\

        Double FM & \textbf{ 11.16}   &     7.27 &    9.91   \\    

        Single+ FM & 11.20     &    12.07  &       9.28   \\ 

        NAS-FM &\textbf{11.16}   &\textbf{6.02}    &\textbf{8.24}  \\
        \bottomrule
    \end{tabular}
      \caption{Ablated of search space using 3 oscillators }
    \label{abt}
\end{table}

\subsection{Ablation Study of Search Space}
The previous methods of FM parameter estimation \cite{chen2022sound2synth} \cite{mitchell2005frequency} \cite{horner1993machine} were usually constrained to a specific FM algorithm. However, our approach puts the different FM algorithms in one search space. 
To demonstrate the significance of the method, we first want to answer whether different timbres have their own FM algorithms that are more suitable for them when operator size is fixed. If the answer is yes, can our method find the algorithm that best fits this timbre?
We conduct our experiments conditioned on three oscillators and enumerate all possible FM algorithms:
\begin{itemize}
\item Nested FM: a carrier with two nested modulators as shown in Fig.~1(b);
\item Formant FM: two carriers sharing a modulator as shown in Fig.~1(c);
\item Double FM: a carrier with two modulators in the same row as shown in Fig.~1(d);
\item Single FM+: a single oscillator adds a single FM. Single FM is shown in Fig.~1(a);
\item NAS-FM (Ours): Constructing a $3*2$ candidate oscillator search space forcing three oscillators discarded to search both FM algorithms and frequency ratios.
\end{itemize}

During the search process, the search space of a fixed FM algorithm equals only contains the frequency ratios of each oscillator. In addition, we set the population size as 30 and max iterations as 20 for each method for a fair   comparison. The results are shown in Table~\ref{abt}.

To answer the first question, we compare the result across all FM algorithms. The Nested FM achieves the best performance on violin and trumpet. However, this algorithm yields the second-worst performance on the Flute. The double FM performs the best on the Flute but worst on the trumpet. Therefore, we find that no fixed FM algorithm achieves the best results for every musical instrument recording. However, our method can always find the best FM configuration for different instruments under the same searching setting. This proves the merit of our well-designed search space.

\subsection{Tuning of Learned FM synthesizers}
In this section, we check the tuning ability of NAS-FM. Two important tasks in sound synthesis are considered: sound morphing and timbre interpolation.
\begin{figure}[!t]
      \centering
      \includegraphics[scale=0.41]{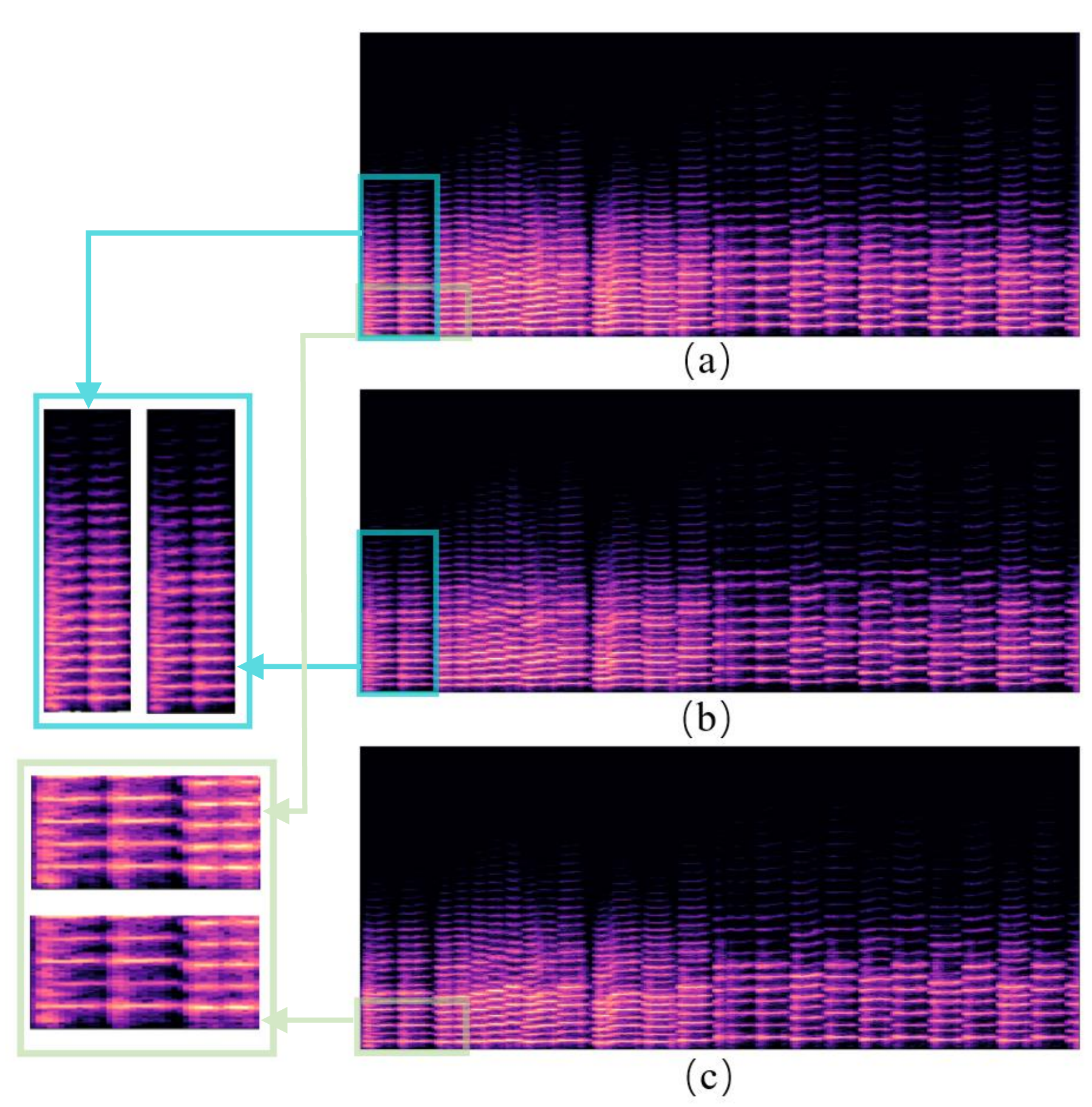}
      \caption{Examples of sound morphing. The top figure is the synthesized sound of the trumpet.  In the middle and bottom figures, we morph the sound by tuning a  certain parameter in our NAS-FM.    }
      \label{sm}
\end{figure}
\subsubsection{Sound Morphing}
Due to the timbre of an instrument being controlled by a set of frequency ratio parameters,   we want to show the ability of our NAS-FM tweaks the timbre from a trained synthesizer by adjusting a few parameters. Begin with a synthesizer designed for the trumpet, Fig.~\ref{sm} (a) is the  synthesized recording of the trumpet using our NAS-FM. More specifically, the searched FM algorithm is the same as the 7-th algorithm  shown in Fig.~\ref{dx7} without the feedback module.
It consists of two parts. The left part is a single FM in which the carrier frequency ratio is three and the modulator is one. The right part includes four oscillators which consist of a double FM with one more nested modulator.  The frequency ratios among a carrier, double modulators and a nested modulator are 7,1,2 and 1, respectively.
In Fig.~\ref{sm}(b), we modify the second formant position from the 7th harmonic to the 10th harmonic by modifying the frequency ratio of the carrier in the right part from seven to ten. In addition, we decrease the second and fourth harmonic by tuning the frequency ratio of the modulator  in the left part from one to two, as shown in Fig.~\ref{sm}(c). We can find that due to the adjustment of one parameter in NAS-FM, the entire distribution and details of frequency can be easily changed.

\subsubsection{Timbre Interpolation}
\begin{figure}[!t]
      \centering
      \includegraphics[scale=0.42]{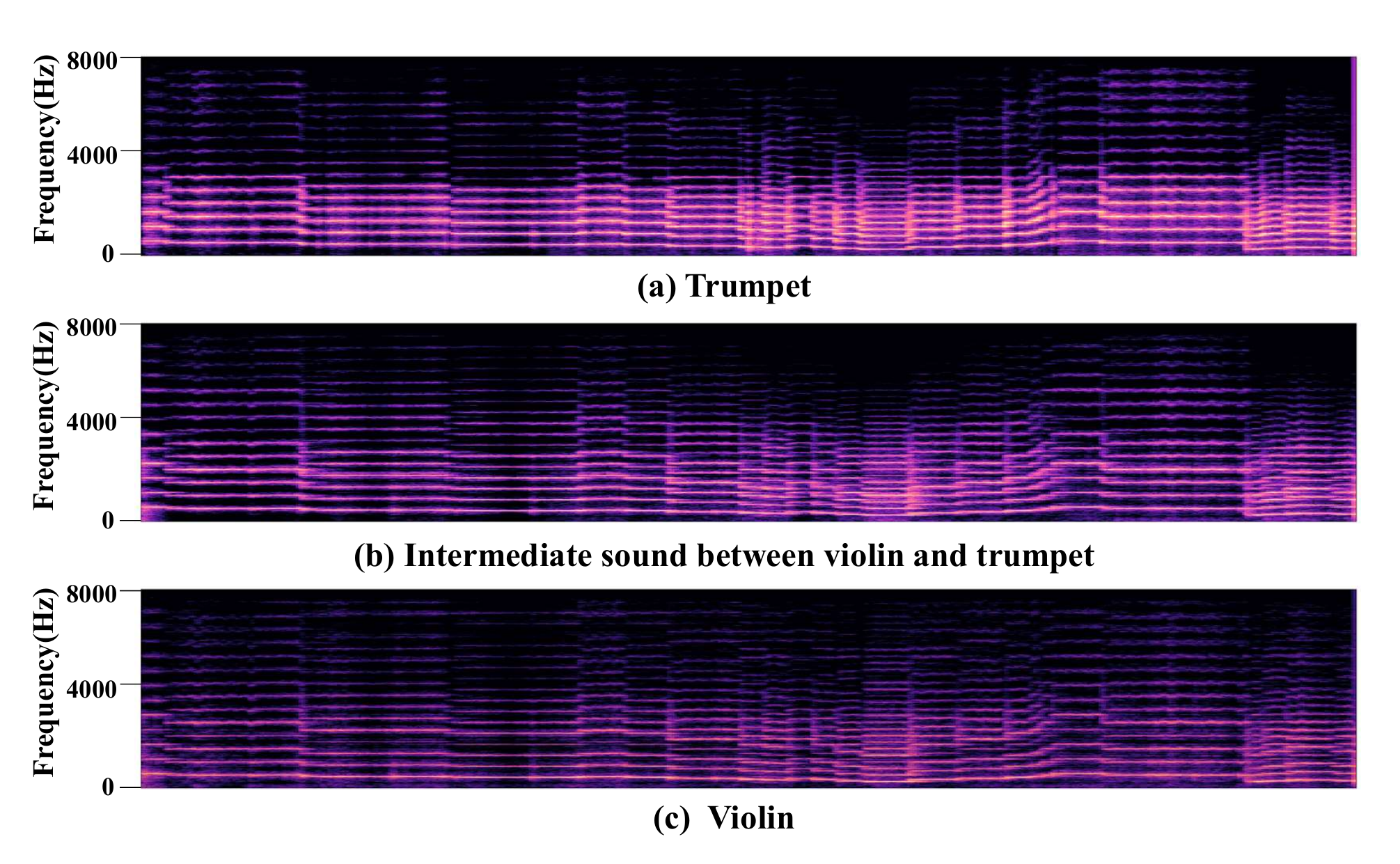}
      \caption{An example of timbre interpolation. The top and bottom figures show the linear spectrogram  of the synthesized sound of the trumpet and violin, respectively. The middle figure shows the interpolation result between the violin and the trumpet.}
      \label{si}
\end{figure}

Another meaningful application is timbre interpolation. Since our approach   determines the FM configuration by searching for the  Fréchet Audio Distance to the target audio clip. We find that the target audio clips can produce new timbre. Therefore, 
we can change our fitness score   to a variant form. For instance, we set the FAD from synthesized audio to violin as $d_v$  and the FAD from synthesized audio to trumpet as $d_t$. Next, we define the novel fitness function as ${d_v} + {d_a} + \left| {{d_v} - {d_a}} \right|$ aiming to find a synthesizer to generate sound regarded as an intermediate timbre between violin and trumpet.
Surprisingly, we use the  supernet trained from violin recordings to conduct the evolutionary algorithm and get an interesting result. As shown in Fig~\ref{si}, we obtain a new instrument recording similar to both violin and trumpet.

\section{Conclusion}
We present NAS-FM, a tunable and interpretable sound synthesizer based on frequency modulation. Given a target sound, we prove that our method can automatically design an FM synthesizer instead of spending a huge time designing it manually.  Meanwhile, our auto-designed synthesizer can achieve comparable results to the handcrafted one. 
Furthermore, Our NAS-FM leverages the widely-used FM synthesizer as the main component in our framework. This makes our method can be directly understood by musicians that can be used to create new sounds without extra knowledge of the neural networks.  

\section*{Acknowledgments}
The research was supported by the Theme-based Research Scheme (T45-205/21-N) and Early Career Scheme (ECS-HKUST22201322), Research Grants Council of Hong Kong.

\bibliographystyle{named}
\bibliography{ijcai23}

\end{document}